\documentclass{aa}
\usepackage{natbib,graphicx}
\usepackage{amssymb, amsmath}
\usepackage[varg]{txfonts}
\begin{document}

\title{The radial distribution of core-collapse supernovae\\
in spiral host galaxies}

\author{A. A. Hakobyan\inst{1}
\and G. A. Mamon\inst{2,3}
\and  A. R. Petrosian\inst{1}
\and D. Kunth\inst{2}
\and M. Turatto\inst{4}
}
\institute{
Byurakan Astrophysical Observatory (BAO) and Isaac Newton Institute (INI) of Chile, Armenian Branch,
Byurakan 0213, Aragatzotn province, Armenia\\
\email{hakobyan@bao.sci.am}
\and
Institut d'Astrophysique de Paris (UMR 7095: CNRS \& Universit\'e Pierre
et Marie Curie), 98bis Bd Arago, 75014 Paris, France
\and
Astrophysics \& BIPAC, University of Oxford, Keble Road, Oxford OX1 3RH, UK
\and
INAF, Osservatorio Astrofisico  di Catania, via Santa Sofia 78, 95123 Catania, Italy
}

\date{Received 30 June 2009 / Accepted 9 October 2009}

\abstract
{}
{With the goal of providing constraints on the nature of the progenitors of
core-collapse (CC) supernovae (SNe),
we compare their radial distribution within their spiral host galaxies with
the distributions of stars and ionized gas in the spiral disks.}
%
{SNe positions are taken from the Asiago catalog for a well-defined sample of
  224 SNe
within 204 host galaxies. The SN radial distances are estimated from the deprojected
separations from the host galaxy nuclei, and normalized both to
the $25^{\rm th} {\rm mag~arcsec}^{-2}$ blue-band isophotal radius and (for the first time) to the
statistically-estimated disk scale length.}
%
{The normalized radial distribution of all CCSNe is consistent with an
exponential law, as previously found, with a possible depletion of CCSNe within
one-fifth of the isophotal radius (not seen with scale-length normalization).
There are no signs of truncation of the exponential distribution of CCSNe out to 7
disk scale lengths.
The scale length of the  distribution of type~II SNe
appears to be significantly larger than that of the stellar disks of their host
galaxies, but consistent with the scale lengths of Freeman disks.
SNe~Ib/c have a significantly smaller scale length
than SNe~II, with little difference between types Ib and Ic.
The radial distribution
of type Ib/c SNe is more centrally concentrated than that of the stars in a
Freeman disk, but is similar to the stellar disk distribution
that we infer for the host galaxies.
All CCSN subsamples are consistent with the
still uncertain distribution of \ion{H}{ii} regions.
The scale length of the CCSN radial distribution shows no
significant correlation with the host galaxy
morphological type, or the presence of bars. However, low luminosity as well as
inclined hosts have a less concentrated distribution (with the statistical
scale-length normalized radial distances) of CCSNe, which are probably a
consequence of metallicity and selection effects, respectively.}
%
{The exponential distribution of CCSNe shows a scale length
consistent with that of the ionized gas confirming the generally accepted hypothesis
that the progenitors of these SNe are young massive stars.
Given the lack of correlation of the normalized radial distances of CCSNe with
the morphological type of the host galaxy, we conclude that
the more concentrated distribution of SNe~Ib/c relative to SNe~II must arise from the higher
metallicity of their progenitors or possibly from a shallower initial mass
function in the inner regions of spirals.}


\maketitle

\section{Introduction}

Most SNe can be assigned to two main physical classes \citep[e.g.,][]{Turatto03,Tur+07}: the gravitational
collapse of young massive stellar cores (Types Ib, Ic and II SNe) and the
thermonuclear explosions of a white dwarf in close binary systems (Type Ia
SNe). While SNe with no or weak hydrogen lines were classified
into type I, it is now understood that there are actually three
spectroscopically and photometrically
distinct subclasses of SNe~I. Type Ia SNe are characterized by spectra with no
hydrogen lines and strong \ion{Si}{II} lines \citep[e.g.,][]{Leibundgut2000,Liv01}.
These SNe appear in galaxies of all morphological types
\citep[e.g.,][]{BBCT99,vdBLF05}.
Type Ib SNe are characterized by spectra with no
evident hydrogen lines, weak or absent \ion{Si}{II} and strong \ion{He}{I} lines. The third
subclass, type Ic SNe, discovered later, shows weak or absent hydrogen, helium
lines, and \ion{Si}{II}. Type II SNe are characterized by the obvious presence of
hydrogen lines. This SN type displays a wide variety of properties:
type II Plateau SNe (SNe~IIP) with flat light curves in the first few months;
type II Linear SNe (SNe~IIL) with a rapid, steady decline in the same period;
the narrow-lined SNe (SNe~IIn), dominated by emission lines with narrow components,
a sign of energetic interaction between the SN ejecta and the circumstellar material;
and type IIb SNe (SNe~IIb), a transitional type, which have early time spectra
similar to SNe~II and late time spectra similar
to SNe~Ib/c \citep[e.g.,][]{Filippenko+90,Filippenko+94}.

With the exception of SNe~Ia, all types of SNe
are rare in early-type galaxies \citep[e.g.,][]{vdBLF05}.
Surprisingly, among morphologically classified hosts of SNe~Ib/c and SNe~II \citet{HPM+2008}
found 22 cases where the host has been classified as an Elliptical or S0 galaxy.
However, all the early-type hosts of SNe~Ib/c and SNe~II
display independent indicators of
recent star formation due to merging or gravitational interaction.

According to theoretical models, the progenitors of SNe~Ib/c are
massive O type stars that have lost most
or all of their hydrogen (and perhaps their helium) envelopes, either by strong winds
as in Wolf-Rayet stars \citep[e.g.,][]{Eldridge+04}, or through the
transfer of material to a binary companion via Roche lobe overflow \citep[e.g.,][]{Heger+03}.
Progenitors of SNe~II are massive stars that retain their hydrogen
envelopes \citep[e.g.,][]{WW86,Hamuy03rev}.

Many observational studies confirm these theoretical
interpretations of
these classes of core-collapse SNe (CCSNe).
The rates of all CCSN types depend on the morphology of the
host galaxies. The rate of SNe~Ib/c and II per unit stellar mass increases by
factors of 3 and 5 respectively from early- to late-type spirals host
galaxies \citep[e.g.,][]{Mannucci+05}. Given the short lifetime of their
massive progenitors, the rate of CCSNe in host galaxies directly traces the
current star formation rate \citep[e.g.,][]{Mal03}. Conversely, when the
star formation rate is known, it can be used to verify the consistency
of the progenitor scenario \citep[e.g.,][]{HB06}.

The observation of progenitor stars of CCSNe in archival pre-explosion images
provides a direct test of the theoretical predictions. Since
massive evolved stars are the most luminous objects in a galaxy, the
progenitors of CCSNe should be directly detectable on pre-explosion
images of nearby host galaxies. The advent of data archives of large telescopes with
high image quality, most importantly that of the
Hubble Space Telescope, has allowed to extend the progenitor detection to
distances larger than 20 Mpc \citep[e.g.,][]{vD+03,Mau+05}.
The observations have generally confirmed  the theoretical predictions mentioned above
\citep[e.g.,][]{Crockett+07,Pastorello+07,Smartt+08}.

The spatial distribution of SNe in host galaxies provides another strong constraint on the nature of SN progenitors.
Various studies show that CCSNe are tightly connected to the disk \citep[e.g.,][]{JM63}
and to the spiral arm structure (\citeauthor{JM63}; \citealp{MvdB76,BTF94,vDHF96}),
which dramatically differs from the SN~Ia distribution \citep[e.g.,][]{FoSch08}. In addition, CCSNe are well
associated with star-forming sites: OB-associations, and \ion{H}{ii} regions
\citep{BTF94,vDHF96,TBP01}.
SNe~Ib/c show a higher degree of association with \ion{H}{ii} regions
than those of type II SNe, suggesting that they arise from a higher mass
progenitor population than the SNe~II \citep{AndeJa08}.
Most recently, \citet{AndeJa09} found that the radial positions of a sample of 177 CCSNe closely follow the
radial distribution of H$\alpha$ emission, implying that these SNe are excellent tracers of star formation
within galaxies.

One can better constrain the nature of the progenitors of CCSNe by
considering quantitative measures of
their radial distribution within their host galaxies,
after correcting for the inclination of the disk.
In their pioneering study, \citeauthor{JM63} found a rapidly decreasing
surface density distribution of SNe, except for an important lack of SNe in the
central regions of spiral galaxies.
The surface density of SNe is known to be exponential \citep{BCC75,VZ77},
although two studies \citep{IK75,GKK80} suggested a ring like distribution.
\citeauthor{BCC75} quantified the
exponential to have a slope that amounts to a scale length of $0.46\pm0.03$
times the optical radius, while \citeauthor{VZ77} found
a scale length of $0.12\pm0.01$ times the optical radius.
The difference in scale length arises from the different definitions of
optical radius.
All SN types, even the SNe Ia, were included in these early studies.
Using a large sample of CCSNe (of known types), i.e. 74 SNe~Ib/c and SNe~II,
\citet{vdB97} found  a scale length of $0.22\,R_{25}$
(with no error bar given), where  $R_{25}$ is the isophotal radius for the
blue-band surface brightness of $\mu_B=25\,\rm mag\,arcsec^{-2}$.
Also, \citet{Bar+92} found that their 99 SNe~II have an exponential
distribution with scale length that amounts to $(0.27\pm0.08)\,R_{25}$,
extending far beyond the optical radius of galaxies.
Moreover, \citeauthor{Bar+92} noticed a sharp decrease in slope of
both SNe~II and SNe~Ib at $1.4\,R_{25}$.

It has rapidly become apparent that SNe~Ib/c are more centrally concentrated
within galaxy disks than are SNe~II.
\citeauthor{Bar+92} noticed that their 22 SNe~Ib show a non-exponential
distribution that increases slope with radius so that the scale
length varies from $(0.15\pm0.02)\,R_{25}$ inside
$0.5\,R_{25}$ to $(0.10\pm0.02)\,R_{25}$ between 0.5 and $1.0\,R_{25}$.
The more centrally concentrated distribution of SNe~Ib/c in comparison to that
of SNe~II was also noticed by \citet{vdB97} and \citet{Wang+97},
but, in contrast to the study of \citeauthor{Bar+92},
the difference with the distribution of SNe~II was not statistically
significant.

Finally, \citet{TPB04} found that, while in the central regions,
SNe~Ib/c are more concentrated than are the
SNe~II, in the outer regions the two distributions are similar.
This result is at odds with the results of \citet{Bar+92}, who found
increasingly dissimilar slopes for the surface densities of SNe~II and of
SNe~Ib/c in the outer regions of spirals.

Additional insight is obtained by comparing the radial distribution of CCSNe
in galaxies of different activity or environment.
\citet{PT90} found that their sample of 8 SNe~II and SNe~Ib within galaxies
hosting AGN were significantly more radially concentrated in
their galaxy hosts than analogous CCSNe in galaxies without active nuclei.
\citet{Petr+05} studying a sample of 12 SNe~II and SNe~Ib/c
in galaxies hosting AGN,  confirmed the above result and found that
SNe~Ib/c in active/star-forming galaxies are more centrally concentrated
than are the SNe~II, but given the small sample, this difference was not
statistically significant. These results were confirmed with larger samples
of CCSNe by \cite{Hak08}, who used
both one-dimensional and multivariate statistics.

The locations of SN explosions in multiple systems have also been studied.
In interacting galaxies, CCSNe are not preferentially located towards the
companion galaxy \citep{Petr+95,Nav+01}. Similarly, the
azimuthal distributions inside the host members of galaxy groups are consistent
with being isotropic (\citealp{Nav+01}).

In this paper, we make use of a considerably larger sample of CCSNe to
reanalyze the distribution of deprojected SN radii normalized both to $R_{25}$ and, for the first time,
to indirect estimates of individual stellar disk scale lengths.
We make statistical comparisons between the radial distributions of CCSNe and
the distribution of blue light and \ion{H}{ii} regions in the disks of spiral galaxies.

The plan of the paper is as follows. The samples of CCSNe and host galaxies
are presented in Sect.~\ref{sample}. We discuss the normalizations in Sect.~\ref{norm}.
The results, given in Sect.~\ref{results}, are discussed and summarized in
Sect.~\ref{discus}. Throughout this article we have assumed a value of
$H_0=75 \,\rm km \,s^{-1} \,Mpc^{-1}$ for the Hubble constant.

\section{The CCSN and host galaxy sample}
\label{sample}

The present investigation is based upon the Asiago Supernova Catalogue (ASC)
\citep{BBCT99}\footnote{http://web.oapd.inaf.it/supern/cat}, updated on
30~September 2008.
This version contains 4730 SNe (86 SNe~I,
2196 SNe~Ia, 304 SNe~Ib/c, 1141 SNe~II and 1003 unclassified SNe) and data for
their host galaxies.
Note that for SNe~II it includes also subtypes IIP, IIL, IIn, IIb.
SNe classified as type~I or type~Ia, or for which the classification was uncertain
(marked as : and ?) or SNe classified  from the light curve only
(labelled as *) were excluded from the present study.
The last SN included in the present investigation is SN~2008fq,
discovered on 15~September 2008.

The analysis of the radial distribution of SNe in
host galaxies requires a  well-defined sample.
We have selected all CCSNe and their hosts with the following criteria:
\begin{enumerate}
\item existing SN spectroscopic classification;
\item host galaxy with morphological type between Sa and Sd,
      excluding peculiar morphologies;
\item host galaxy inclination $i \leq 50^\circ$ to minimize absorption and
      projection effects ($0^\circ$ means face-on);
\item known position angle of the major axis of the host galaxy;
\item known host galaxy heliocentric radial velocity;
\item known integrated $B$ magnitude of the host galaxy;
\item known position of the SN with respect to the galaxy center.
\end{enumerate}

These criteria led to the selection of 239 CCSNe within 216 host galaxies, of which 61 are of type Ib/c
(in 61 hosts) and 178 are of type II (in 162 hosts).
Our sample includes galaxies with multiple SNe, in particular 19 galaxies with 2 SNe and 1 with 5 SNe.
In order to use homogeneous, updated  information
for all these galaxies, we re-extracted all the main galaxy data from
HyperLeda\footnote{http://leda.univ-lyon1.fr} \citep{Paturel+03}
such as absolute magnitudes and isophotal diameters,
both corrected for Galactic and host galaxy internal extinction
(\citealp{SFD98,BGPT95}, respectively)
and the former corrected for Virgocentric infall \citep{TPE02}.

The distribution of
morphological types of the 216 host galaxies is peaked around Sbc and Sc.
130 host galaxies are not  barred with
38 SNe of type Ib/c and 105 SNe of type II. 86 hosts are barred and
have 23 SNe of type Ib/c and 73 SNe of type II.

Because of projection effects, some of the SNe showing small
distances to the center may lie quite far away from the center.
With the assumption that the CCSNe have young progenitors
that are located in the disks of the galaxies, a more realistic separation
may be derived, using the inclinations of the host galaxies.

\begin{figure}[ht]
\centering
\includegraphics[width=\hsize]{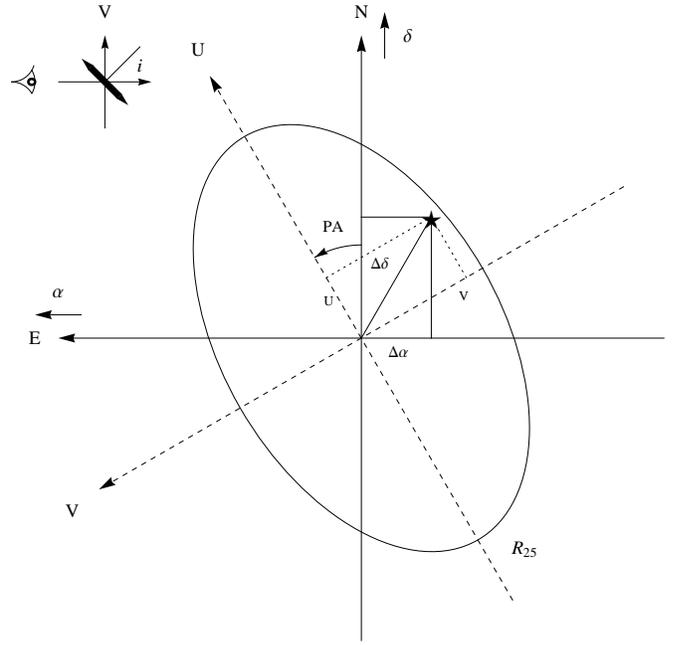}
\caption{Location of the SN within its host galaxy. The
center of the galaxy is at the origin of coordinate systems
and the \emph{star symbol} is the projected location of the SN.
$\Delta \alpha$ and $\Delta \delta$ are offsets of the SN in equatorial coordinate
system, \textsc{u} and \textsc{v} are coordinates of the SN in host galaxy coordinate system
along the major (U) and the minor (V) axis, respectively.
The \emph{inset in the upper-left corner} illustrates the inclination of
the polar axis of the galaxy with respect to the line of sight.
\label{snloc}}
\end{figure}
Fig.~\ref{snloc} illustrates the
geometrical location of a SN in the plane of the host galaxy.
The coordinates of the SN in its host galaxy coordinate system are
\begin{equation}
\begin{split}
 \textsc{u} = \Delta \alpha \,\sin {\rm PA} + \Delta \delta \,\cos {\rm PA} \ ,\\
 \textsc{v} = \Delta \alpha \,\cos {\rm PA} - \Delta \delta \,\sin {\rm PA} \ , \\
\end{split}
\label{uvdefs}
\end{equation}
and the true radial distance of the SN to the galactic center satisfies
\begin{equation}
R_{\rm SN}^2 = {\textsc{u}}^2+\left ({{\textsc{v}}\over \cos i} \right)^2 \ .\\
\label{RSN}
\end{equation}

\section{Radius normalization}
\label{norm}

Following \citet{PT90}, \citet{Bar+92} and \citet{vdB97},
we first normalize the SN radius to the $25^{\rm th}$ magnitude isophotal radius,
$R_{25}$ (corrected for dust extinction).

The radial surface brightness distribution of spiral disks are well
described by an exponential law \citep[e.g.,][]{Free70}
$\Sigma^d(R)=\Sigma_0^d \exp(-R/h)$,
where $R$ is radius measured along the disk from the center,
$h$ is the scale length of the exponential disk,
and $\Sigma_0^d$ is the central surface brightness.
In terms of $r=R/R_{25}$, the surface brightness distribution of disks is
$\Sigma^d(r)=\Sigma_0^d \exp(-r/\tilde{h})$, where the scale length follows
the relation
\begin{equation}
\tilde{h} = {h\over R_{25}} = {2.5/\ln 10\over 25-\mu_0^{\rm disk}} \ , \\
\label{mju}
\end{equation}
where $\mu_0^{\rm disk}$ is the \emph{disk} central $B$-band surface brightness.

\citet{Free70} asserted that the extrapolated disk central surface blue brightness,
corrected for inclination and Galactic absorption,
is almost constant and equal to $\mu_0^{\rm Freeman}=21.65\pm0.30$ $\rm mag \, arcsec^{-2}$.
Inserting $\mu_0^{\rm Freeman}$ into equation~(\ref{mju}),
we obtain the stellar disk scale length of $\tilde{h}_{\rm Freeman}=0.32\pm0.03$.
Hereafter, we denote this stellar distribution as the \emph{Freeman disk}.

However, the Freeman disk is an oversimplified model:
the central disk surface
brightness is fainter for late-type spirals, and also decreases with
increasing  scale length $h$, as found by \citet{GdB01} in the $B$ and $R$
bands, as well as \citet{Graham01} in the $K$ band.

Therefore, rather than assuming that the surface density profile of SNe has a
characteristic scale that is proportional to the isophotal radius of their
host galaxy, as above, we can alternatively assume  that this SN scale is proportional
to the stellar disk scale length of their host galaxy, and thus normalize the
SN galactocentric radii to the scale lengths of the disks of their host galaxies.

Ideally, we could perform a bulge/disk decomposition for each of our galaxies
to obtain the individual disk scale length. However, this is beyond the scope of the
present paper, and we have chosen instead to adopt a statistical approach, by
relying on the fits of $\log h$ as a function of disk
absolute magnitude $M_{\rm disk}$ by \citet{GW08}.
Noting that the slope in the $\log h$ vs. $M_{\rm disk}$ relation of types Sc and earlier
depends little on the waveband (Table~9 of \citeauthor{GW08}), we assume that
the same holds for later morphological types, and thus take the slopes in
the $B$ band from those established by \citeauthor{GW08} (their Table~10)
in the $K$ band for
each morphological type. We also assume that the difference
in $\log h$ normalizations between the $B$ and $K$ bands is independent of
morphological type. This finally gives us
\begin{equation}
\log h = a - 0.4\, b \left (M_{\rm disk}+20\right) \ ,
\label{heq}
\end{equation}
where $h$ is in kpc and $M_{\rm disk}$ is the disk absolute blue magnitude.
The values for the normalization $a$ and slope $b$ are given in
Table~\ref{hparstab}. Note that \citeauthor{GW08} obtained their parameters
after correcting for extinction.

\begin{table}[ht]
\begin{center}
\caption{Parameters of the scale length vs. luminosity relation (eq.~[\ref{heq}]).
\label{hparstab}}
\begin{tabular}{lcc}
\hline
\hline
Morphological type & $a$ & $b$ \\
\hline
Sa & 0.35 & 0.70 \\
Sb & 0.39 & 0.58 \\
Sc & 0.42 & 0.39 \\
Sd & 0.46 & 0.23 \\
\hline
\end{tabular}
\end{center}
\end{table}

We estimate the disk magnitudes by adopting the median
$B$ band bulge-to-disk ratios $\log({\rm B/D})$ from Fig.~7 of \citet{GW08}, i.e.
$M_{\rm disk} = M +2.5\,\log \left (1+ {\rm B/D} \right )$.

\begin{figure}[t]
\centering
\includegraphics[width=\hsize]{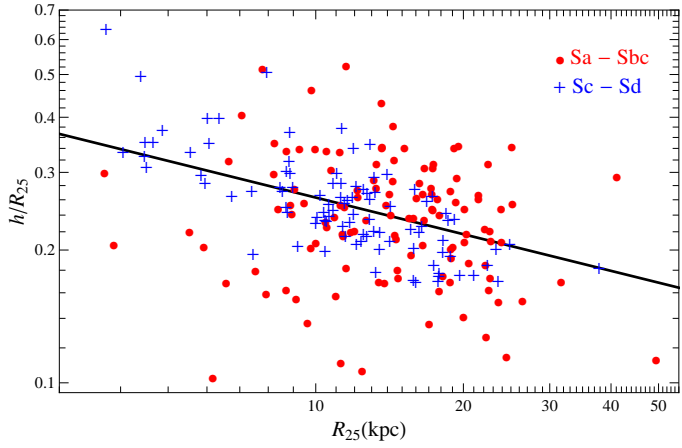}
\caption{Comparison of $h / R_{25}$ and $R_{25}$, where the scale
length $h$ is computed according to equation~(\ref{heq}) and
Table~\ref{hparstab}.
\emph{Red filled circles} and \emph{blue crosses} respectively show the
early-type (Sa-Sbc) and late-type (Sc-Sd) galaxies.
The \emph{solid line} is a best fit to the full sample of CCSNe.}
\label{RsnR25h}
\end{figure}

Fig.~\ref{RsnR25h} illustrates the comparison of the two normalized radius
indicators. Also shown is the best fit line
\begin{equation}
 \log \left ({h \over R_{\rm 25}}\right ) = (-0.30\pm0.04) +
 (-0.27\pm0.04) \, \log \left ({R_{25}\over {\rm kpc}}\right) \ .
\label{RSNoverhfit}
\end{equation}
The galaxies with large isophotal radii have a smaller ratio of disk scale
length to isophotal radius, suggesting (eq.~[\ref{mju}]) that large spiral
galaxies
have higher central disk surface brightness, as found by
\citet{Driver+06}. Moreover, since
the central disk surface brightness can be written
\begin{equation}
\mu_0^{\rm disk} = M_{\rm disk} + 5\,\log h + 38.74 \ ,
\label{mu01}
\end{equation}
where $h$ is in kpc, and combining equations~(\ref{heq}) and (\ref{mu01}),
one finds that
$\mu_0^{\rm disk} = 2.5\,b/(2\,b-1)\,\log h + constant$, i.e. higher central surface
brightness for larger scale length when $b<1/2$, which, according to
\citeauthor{GW08}, occurs for late-type
spirals (see Table~\ref{hparstab} above).

On average, we find $\langle h /R_{25} \rangle = 0.26\pm0.02$ (where the errors here
and below are on the mean).
Using equation~(\ref{mju}), one deduces $\mu_0^{\rm disk} = 20.82\pm0.30$,
i.e. about 0.8 magnitude brighter than the Freeman disk.

\begin{figure}[ht]
\centering
\includegraphics[width=0.8\hsize]{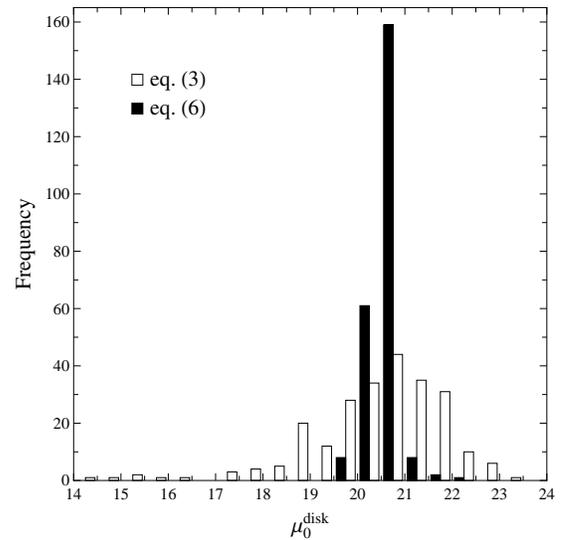}
\caption{Distribution of central disk surface brightness obtained from
equations~(\ref{mju}) (\emph{open bars}) and (\ref{mu01}) (\emph{filled
  bars}),
using our statistical measurement of disk scale length.
\label{mu0dist}}
\end{figure}

From a statistical point of view the central disk surface brightness
for the galaxies of our sample can be obtained in either two ways:
\begin{enumerate}
\item Given the galaxy isophotal radius $R_{25}$ and \citeauthor{GW08}'s
statistical estimate of $h$, using equation~(\ref{mju});
\item Given the galaxy absolute magnitude $M$ and
\citeauthor{GW08}'s statistical estimates of $M_{\rm disk}-M$
(given the galaxy type) and $h$
(given the galaxy type and disk luminosity),
using equation~(\ref{mu01}).
\end{enumerate}
The first method yields
$\mu_0^{\rm disk}=20.91^{+1.36}_{-1.00}$, 3/4
of a magnitude  brighter than the Freeman disk,
with some scatter, while the second method yields
$\mu_0^{\rm disk}=20.79^{+0.15}_{-0.26}$, which is even brighter, but with very little scatter.
The distribution of the central disk surface brightness is given in
Fig.~\ref{mu0dist}.

Which measure of $\mu_0^{\rm disk}$ is more accurate?
The uncertainty on the first measure of $\mu_0^{\rm disk}$ (eq.~[\ref{mju}]) is
\[
\sigma_{\mu_0^{\rm disk}}^2 = \left ({2.5\over \ln 10}\,{R_{25}\over h}\right )^2\,
\left [\left ({\sigma_h\over h}\right )^2 + \left ({\sigma_{R_{25}}\over
    R_{25}} \right )^2 \right ] \ .
\]
Using equation~(\ref{mu01}), the uncertainty on the second measure
of $\mu_0^{\rm disk}$ is
\[
\sigma_{\mu_0^{\rm disk}}^2 = \left ({5\over \ln 10}\right )^2\,\left ({\sigma_h\over
  h}\right )^2 + \sigma_{M_{\rm disk}-M}^2 + \sigma_M^2 \ .
\]
Hence the difference in the square uncertainties
on the first (eq.~[\ref{mju}]) and second (eq.~[\ref{mu01}]) measures of
$\mu_0^{\rm disk}$ is equal to
\[
\left ({2.5\over \ln 10}\right)^2
\left[\left ({R_{25}\over h}\right  )^2\!-\!4\right]
\left ({\sigma_h\over  h}\right )^2
+ \left ({2.5\over \ln 10}\right)^2\left ({\sigma_{R_{25}}\over R_{25}}
\right )^2  -  \sigma_{M_{\rm disk}-M}^2 - \sigma_M^2\,.
\]
Given that, on average, $R_{25}/h \simeq 3.6$, i.e. $(R_{25}/h)^2 = 12.8$,
and given typical uncertainties $\sigma_h/h \approx 0.25$ (from Fig.~10a of
\citeauthor{GW08}), $\sigma_{M_{\rm disk}-M} = 2.5 ({\rm B/D})/(1+{\rm B/D})\,
\sigma \log ({\rm B/D}) \la
0.25$ (from Fig.~6 of \citeauthor{GW08}), then even with negligible errors on
$R_{25}$ the uncertainty on the first measure of $\mu_0^{\rm disk}$ (eq.~[\ref{mju}]) is
considerably greater than that of the second measure of $\mu_0^{\rm disk}$
(eq.~[\ref{mu01}]).
We therefore adopt equation~(\ref{mu01}) to estimate $\mu_0^{\rm disk}$.

In comparison, repeating the same analysis for the 1500 galaxies in a
reference sample found in HyperLeda with
the same selection as in points 2 to 6 of Sect.~\ref{sample} (plus a limit on
redshift), we find (with eq.~[\ref{mu01}]) a mean central disk
surface brightness of $\left\langle \mu_0^{\rm disk} \right\rangle = 20.78$,
very similar to the central surface brightness inferred from
equation~(\ref{mu01}),
indicating that SN hosts do not have centrally brighter disks than
other galaxies selected in the same way.

\section{Data analysis and results}
\label{results}

\subsection{Isophotal radius normalization}
\label{resisonorm}

\begin{figure}[ht]
\centering
\includegraphics[width=\hsize]{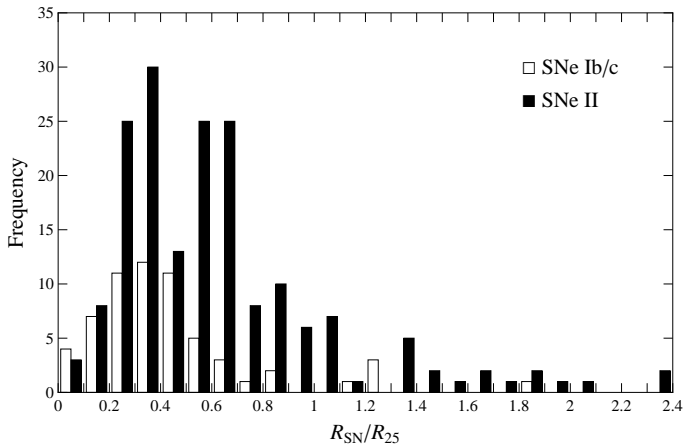}
\caption{Distribution of deprojected isophotal-normalized radial distances
  for SNe of types Ib/c (\emph{open bars}) and II (\emph{filled bars}).}
\label{his1}
\end{figure}

Fig.~\ref{his1} presents the histograms of the relative radial distributions
of types Ib/c and II SNe in the sample of spiral host galaxies.
45 out of 61 SNe~Ib/c (74\%) are found to be
located within relative distance $R_{\rm SN}/R_{25}=0.5$ from  the nuclei of
their host galaxies, compared to 79 out of 178 (44\%) of all SNe~II.
The average relative radial distances
for types SNe~Ib/c, SNe~II, and all CCSNe are
$0.45\pm0.04$ ($N_{\rm SN}=61$), $0.62\pm0.03$ ($N_{\rm SN}=178$) and
$0.58\pm0.03$ ($N_{\rm SN}=239$), respectively.

The observational deficit of SNe
in the central regions of remote galaxies relative to nearer
galaxies, the \cite{Shaw79} effect, is found to be
more important for deep photographic SN searches and negligible
for visual and CCD searches in nearby galaxies \citep[e.g.,][]{CTTB+97}.
The relative lack of CCSNe at the largest relative galactocentric distances
for nearby
galaxies might be caused by the limited fields of view of
various SN search programs.

\begin{figure}[ht]
\centering
\includegraphics[width=\hsize]{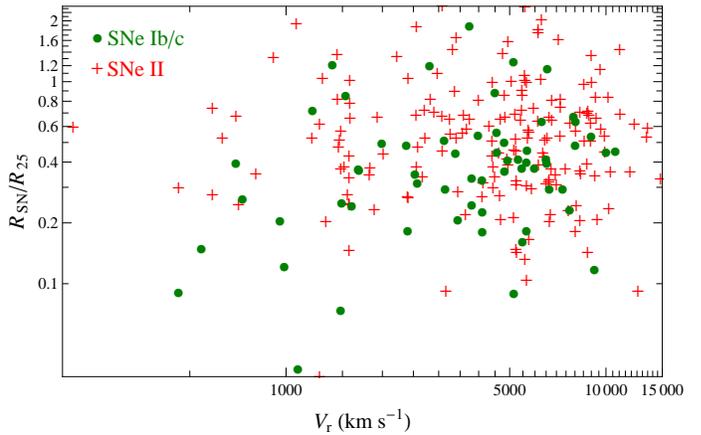}
\caption{Deprojected isophotal-normalized SN-to-host distances
vs. radial velocity of host galaxy (as a proxy for distance).
SNe~Ib/c and SNe~II are shown as \emph{green filled circles} and
\emph{red plus signs}, respectively.}
\label{RsnR25V}
\end{figure}

Fig.~\ref{RsnR25V}, similar to Fig.~1 of
\cite{Wang+97}, illustrates the relative
distance of CCSNe from the centers of their hosts
vs. the radial velocity of the host galaxy.
Spearman rank correlation tests indicate that SNe~II show no trend for
$R_{\rm SN}/R_{25}$ vs. host galaxy radial velocity (rank correlation
coefficient $r_{\rm S} = -0.002$). However, these tests indicate strong
positive trends between galaxy radial velocity and $R_{\rm SN}/h$ ($r_{\rm S}
= 0.13$ with probability $P=0.04$ of a stronger trend occurring by chance) for SNe~II
on one hand, and both $R_{\rm SN}/R_{25}$ ($r_{\rm S}=0.19, P=0.06$) and $R_{\rm
  SN}/h$ ($r_{\rm S}=0.20, P=0.05$) for SNe~Ib/c
on the other hand.
However, if one removes the nearest galaxies ($V_{\rm r} < 1000 \, \rm km \,
s^{-1}$),
the trends become weaker and no longer statistically significant:
$r_{\rm S}=0.10, P=0.10$ for SNe~II with scale length
normalization, and $r_{\rm S}=0.08, P=0.27$ for SNe~Ib/c with either
normalization.

We therefore limit our CCSN sample to distances $V_{\rm r} > 1000 \, \rm km \,
s^{-1}$, thus retaining 224 SNe of the original 239, living in 204 host galaxies.
Table~\ref{tabl1} displays the matrix of SNe type versus host galaxy
morphological type in this reduced sample.
The SNe~Ib/c are equally split into 25 early-type spiral hosts (Sa to Sbc) and 30
late-type (Sc to Sd), while among the SNe~II, 102 are in early-type spirals and only
67 in late-types. But this preference for late-type spirals of SNe~Ib/c is
not statistically significant (from binomial statistics).

\begin{table}[ht]
\begin{center}
  \caption{Distribution of SN types and host galaxy morphological types.}
  \label{tabl1}
  \begin{tabular}{l r r r r r r r r} \hline
  \hline
  &\multicolumn{1}{c}{  Sa} &\multicolumn{1}{c}{  Sab} &\multicolumn{1}{c}{  Sb} &\multicolumn{1}{c}{  Sbc} &\multicolumn{1}{c}{  Sc} &\multicolumn{1}{c}{  Scd} &\multicolumn{1}{c}{  Sd} &\multicolumn{1}{c}{ Total} \\ \hline
  SNe~Ib/c & 2 & 3 & 10 & 10 & 26 & 4 & 0 & 55 \\
  SNe~II & 14 & 7 & 34 & 47 & 38 & 17 & 12 & 169 \\
  $N_{\rm gal}$ & 16 & 10 & 42 & 48 & 57 & 20 & 11 & 204 \\ \hline
\end{tabular}
\end{center}
\noindent
Notes: The bottom row lists the number of unique host galaxies.
\end{table}

\begin{figure*}[ht]
\centering
\includegraphics[width=\hsize]{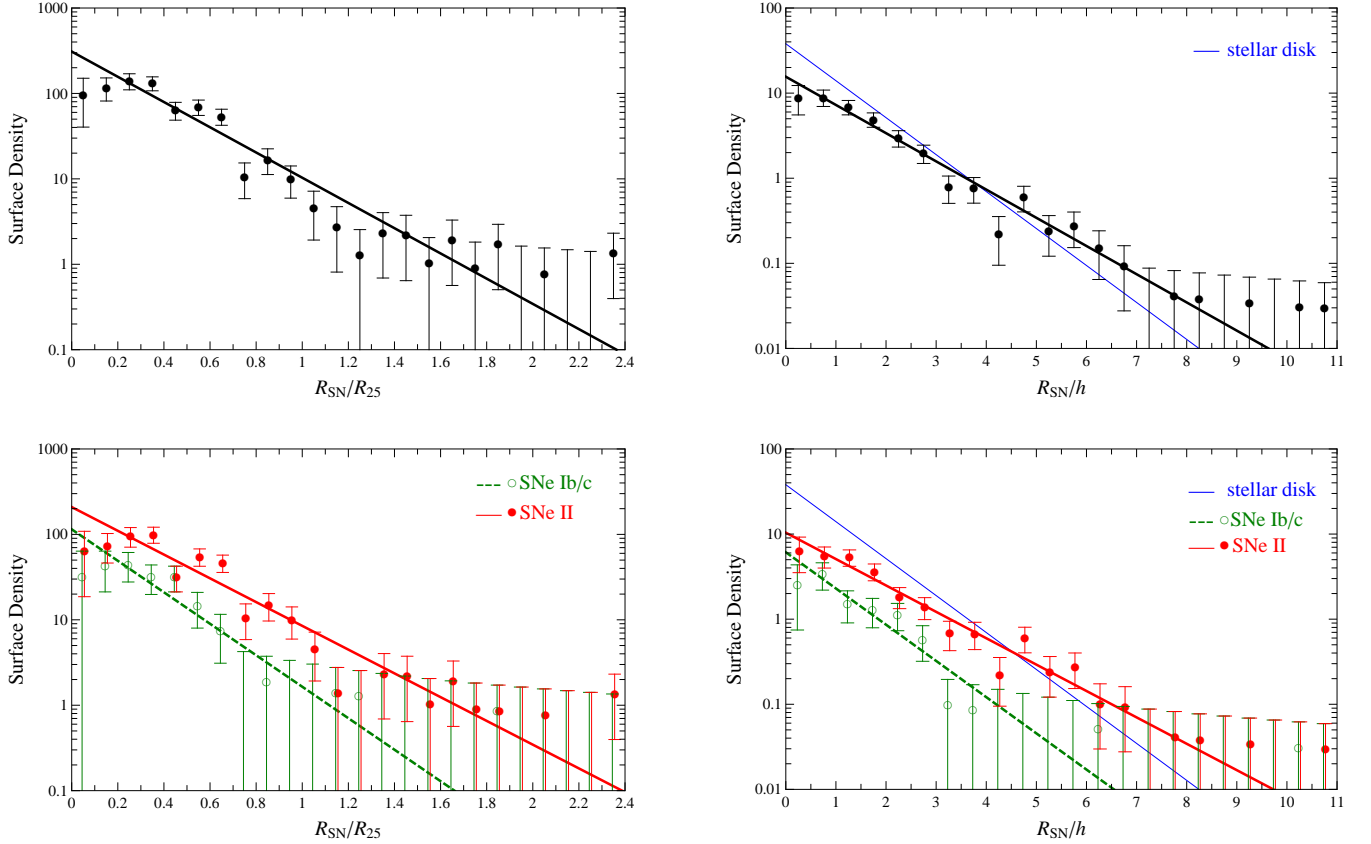}
\caption{\emph{Top}: surface density profiles (with arbitrary normalization)
of all CCSNe, with host galaxy isophotal
radius normalization (\emph{left}), and with disk scale length normalization
(\emph{right}).
The \emph{thick solid line} shows the maximum likelihood exponential surface
density profile of CCSNe.
The \emph{thin solid line} represents the stellar
exponential disk of the host galaxies.
\emph{Bottom}: same as top panels, for SNe~Ib/c
(\emph{green open circles}) and SNe~II (\emph{red filled circles}).
The \emph{thick lines} indicate the maximum likelihood
exponential surface density profiles for the SNe~Ib/c (\emph{green dashed}) and SNe~II
(\emph{red solid}).
The error bars  assume a Poisson distribution
(with $\pm1$ object if none is found).}
\label{sdensall}
\end{figure*}

The distribution of radial distances can be better understood by considering the
normalized surface density profile of SNe within their host
galaxies. We have determined the
surface density of CCSNe, $\Sigma_{j}^{\rm SN} = n_{j} / \pi (r_{j+1}^2 - r_{j}^2)$ where $r_j$
is the relative inner radius of the circular annuli of width $r_{j+1} - r_{j}$ ($j$-th bin),
and $n_{j}$ indicates the number of CCSNe in the $j$-th bin.
The top left panel of Fig.~\ref{sdensall} shows that the stacked surface density
distribution of all 224 CCSNe appears consistent with an exponential law, except for a lack of
SNe in the central regions ($R_{\rm SN} < 0.2\,R_{25}$).

If the surface density of CCSNe is an exponential
function of $\tilde{r}=R_{\rm SN}/R_{25}$
(i.e. $\Sigma^{\rm SN}(\tilde{r})=\Sigma_0^{\rm SN} \exp(-\tilde{r}/\tilde{h}_{\rm SN})$,
where $\tilde{h}_{\rm SN}$ is the scale length
for the distribution on disk, and $\Sigma_0^{\rm SN}$ is the central surface
density of CCSNe), then
the probability that a SN is observed at normalized radius $\tilde{r}_i=R_{\rm SN}^i/R_{25}^i$ is
\begin{equation}
\begin{split}
 p(\tilde{r}_i | \tilde{h}_{\rm SN}) =
 {2\pi \tilde{r}_i\Sigma_0^{\rm SN}\exp\left(-\tilde{r}_i/\tilde{h}_{\rm SN}\right) \over
 \int_{0}^\infty 2\pi \tilde{r} \Sigma_0^{\rm SN}\exp\left(-\tilde{r} /\tilde{h}_{\rm SN}\right) d\tilde{r}} \\
 =  {\tilde{r}_i\over \tilde{h}_{\rm SN}^2}\, \exp\left(-\tilde{r}_i/\tilde{h}_{\rm SN}\right) \ . \\
\end{split}
\label{probr}
\end{equation}

Equation~(\ref{probr}) assumes no truncation in the normalized SN radii, and
indeed we consider all SN, whatever their distance to the center of their
host galaxy. The likelihood of the set of $\{\tilde{r}_i\}$ is
\begin{equation}
{\cal L}(\tilde{h}_{\rm SN})=\prod_{i=1}^{N_{\rm SN}} p(\tilde{r}_i | \tilde{h}_{\rm SN}) \ ,
\end{equation}
and maximizing $\ln {\cal L}$ amounts to maximizing
\[
{\sum_{i=1}^{N_{\rm SN}} \ln \tilde{r}_i} - {1\over{\tilde{h}_{\rm SN}}}
\sum_{i=1}^{N_{\rm SN}} \tilde{r}_i - 2 \,N_{\rm SN} \ln \tilde{h}_{\rm SN}
\]
for $\tilde{h}_{\rm SN}$ (where we made use of eq.~[\ref{probr}]),
which yields
\begin{equation}
\tilde{h}_{\rm SN} = {{\sum_{i=1}^{N_{\rm SN}} \tilde{r}_i}\over {2 \,N_{\rm SN}}} =
      {1\over2}\,\left\langle \tilde{r}_i\right \rangle \ ,
\label{htildeML}
\end{equation}
i.e., the maximum likelihood exponential scale length is half the arithmetic
mean of the radial distances.
The integral of the denominator of equation~(\ref{probr})
yields the normalization
\begin{equation}
\Sigma_0^{\rm SN} = {{N_{\rm SN}}\over {2 \pi \tilde{h}_{\rm SN}^2}} \ .
\end{equation}

We check the goodness of fit using the Kolmogorov-Smirnov (KS) test on the
normalized cumulative distributions of all SN types, where the exponential model has a
cumulative normalized distribution
$E(\tilde{r}) = 1 - (1 + {\tilde{r}/\tilde{h}_{\rm SN}}) \exp (- {\tilde{r}/ \tilde{h}_{\rm SN}})$.

\begin{table}[t]
\begin{center}
  \caption{
Maximum likelihood fits of an exponential surface density profile to the
radial distribution of
CCSNe.}
\label{hMLall}
\tabcolsep 3.3pt
  \begin{tabular}{lrcccc} \hline
    \hline
\multicolumn{1}{c}{Sample}&$N_{\rm SN}$&$\tilde{h}_{\rm SN}$&$\hat{h}_{\rm SN}$&$\tilde{P}_{\rm KS}$&$\hat{P}_{\rm KS}$\\
\multicolumn{1}{c}{(1)}&(2)&(3)&(4)&(5)&(6)\\
\hline
    Full sample&224&$0.29\pm0.01$&$1.25\pm0.06$&0.194&0.646\\
\hline
    Sa-Sbc&127&$0.31\pm0.02$&$1.35\pm0.09$&0.235&0.412\\
    Sc-Sd&97&$0.28\pm0.02$&$1.13\pm0.07$&0.890&0.872\\
    Sa-Sbc (SNe~Ib/c)&25&$0.23\pm0.04$&$0.93\pm0.19$&0.098&0.213\\
    Sc-Sd \,\ (SNe~Ib/c)&30&$0.24\pm0.03$&$1.03\pm0.13$&1.000&0.986\\
    Sa-Sbc (SNe~II)&102&$0.32\pm0.02$&$1.46\pm0.10$&0.441&0.965\\
    Sc-Sd \,\ (SNe~II)&67&$0.29\pm0.02$&$1.17\pm0.09$&0.783&0.793\\
    Sa-Sd \,\ (without bars)&133&$0.29\pm0.02$&$1.18\pm0.08$&0.273&0.707\\
    Sa-Sd \,\ (with bars)&91&$0.30\pm0.02$&$1.37\pm0.10$&0.453&0.931\\
    Sa-Sbc (without bars)&58&$0.32\pm0.03$&$1.26\pm0.14$&\textbf{0.043}&0.523\\
    Sa-Sbc (with bars)&69&$0.29\pm0.02$&$1.43\pm0.13$&0.899&0.973\\
    Sc-Sd \,\ (without bars)&75&$0.26\pm0.02$&$1.11\pm0.09$&0.977&0.997\\
    Sc-Sd \,\ (with bars)&22&$0.32\pm0.04$&$1.19\pm0.14$&0.884&0.829\\
    $-23 < M_{\rm disk} \leq -20.5$&123&$0.27\pm0.02$&$1.06\pm0.07$&0.586&0.933\\
    $-20.5 < M_{\rm disk} < -18$&101&$0.32\pm0.02$&$1.49\pm0.10$&0.308&0.603\\
    $19.65 < \mu_0^{\rm disk} \leq 20.60$&102&$0.28\pm0.02$&$1.35\pm0.10$&0.895&0.840\\
    $20.60 < \mu_0^{\rm disk} < 21.85$&122&$0.30\pm0.02$&$1.18\pm0.08$&0.203&0.803\\
    $0^\circ \leq i \leq 30^\circ$&68&$0.26\pm0.02$&$1.11\pm0.10$&0.912&0.453\\
    $30^\circ < i \leq 50^\circ$&156&$0.31\pm0.02$&$1.32\pm0.08$&0.153&0.719\\
\hline
    SNe~Ib&20&$0.23\pm0.03$&$0.90\pm0.11$&0.765&0.993\\
    SNe~Ic&27&$0.22\pm0.03$&$0.97\pm0.18$&0.671&0.644\\
    SNe~Ib/c&55&$0.24\pm0.02$&$0.99\pm0.11$&0.476&0.707\\
    SNe~II&169&$0.31\pm0.02$&$1.34\pm0.07$&0.328&0.746\\
\hline
\multicolumn{2}{l}{stars (in Freeman disk)}&$0.32\pm0.03$&$1.23\pm0.17$&&\\
\multicolumn{2}{l}{stars (in SN host disks)}&$0.26\pm0.02$&\multicolumn{1}{l}{1.00}&&\\
\multicolumn{2}{l}{\ion{H}{ii} regions (Freeman disk)}&$0.26\pm0.13$&$1.00\pm0.50$&&\\
\multicolumn{2}{l}{\ion{H}{ii} regions (SN host disk)}&$0.21\pm0.11$&$0.80\pm0.40$&&\\
    \hline
  \end{tabular}
\end{center}
\noindent
Notes: Column~1 gives the CCSN sample, Col.~2 number of CCSNe in the sample,
Col.~3 the maximum likelihood value of $\tilde h_{\rm SN} = h_{\rm SN}/R_{25}$,
Col.~4 the maximum likelihood value of $\hat{h}_{\rm SN} = h_{\rm SN}/h$, and
Cols.~5 and 6 give the KS test probabilities that
the surface density distribution of CCSNe is consistent with an exponential
law with the isophotal radius (Col.~5) and scale length (Col.~6)
normalizations.
The four last lines are not for the SNe, but for the stars (first two
of these lines) and \ion{H}{ii} regions (last two lines),
with the
centrally extrapolated surface brightness taken from the Freeman disk  (first and third of these lines)
and the hosts (second and fourth of these lines).
The value of unity in Col.~4 in the second of these lines is a direct consequence of
$h_{\rm SN} = h$ and thus has no uncertainty associated with it.
The values of Col.~4 in the first of these lines is scaled relative to that
of the the second of these lines, according to the mean
value of $\langle h /R_{25} \rangle$ (Sect.~\ref{norm}).
Similarly the last two lines of Col.~4 are scaled to the corresponding values in Col.~3.
The statistically significant deviation from an exponential law is highlighted
in bold.
\end{table}

Table~\ref{hMLall} presents the maximum likelihood fits of the exponential
surface density distributions. One sees that \emph{all SN samples show stacked
surface number density distributions that are consistent with an exponential
distribution}, with the exception of non-barred early-type spiral hosts (among
which are the three galaxies of our 224 SN host galaxy sample with CCSNe
beyond $2\,R_{25}$).

However, Table~\ref{hMLall} indicates that \emph{the distribution of SNe~Ib/c is significantly
more centrally concentrated than the distribution of stars in a Freeman disk.}
More precisely, \emph{SNe Ib/c appear 30\% more centrally
concentrated than SNe~II.}
The scale length of SNe~Ib/c is consistent with the
scale length of stars in the host galaxies, while the SNe~II appear 20\% less
concentrated than the stars in the host galaxies, but as concentrated as the stars in
Freeman disks.
The host galaxy morphological type appears to play no role in the
distribution of SNe. However, the SNe in nearly face-on host galaxies appear
more centrally concentrated than the SNe in more inclined host galaxies.
This indicates that observational selection effects (dust extinction and
confusion) prevent the observation of SNe near the centers of inclined disks.

The bottom left panel of Fig.~\ref{sdensall} highlights the differences
in the radial distributions of  SNe~Ib/c and SNe~II.
While SNe~II have a 30\% larger scale length,
both classes of SN types show an important
(statistically significant for the SNe~II) drop in
the center ($R_{\rm SN}<0.2\,R_{25}$).

The surface density distributions of \ion{H}{ii} regions in spiral galaxy disks can
also be represented reasonably well by an exponential function
\citep[e.g.,][]{HK83,Ath+93,GGAB02}, the exceptions being mainly due to
unsatisfactory fits of the maxima near the center. \citeauthor{Ath+93} found
that the \ion{H}{ii} regions are marginally more concentrated to the centers of spiral galaxies
than are the disk stars, with
a ratio of scale lengths of $h^{\rm \ion{H}{ii}} / h$ is $0.8\pm0.4$.
In units of $R_{25}$, the ionized gas scale length is then
$\tilde{h}_{\rm Freeman}^{\rm \ion{H}{ii}} = 0.26\pm0.13$
if we use the Freeman disk normalization, i.e.
Freeman disk scale length of $\tilde{h}_{\rm Freeman}=0.32\pm0.03$.
If we use the CCSN host disk normalization, i.e.
CCSN host disk average scale length of
$\tilde{h}_{\rm hosts} = \left\langle h/R_{25}\right\rangle = 0.26\pm0.02$,
then $\tilde{h}_{\rm hosts}^{\rm \ion{H}{ii}} = 0.21\pm0.11$.
Given the large uncertainty on the ratio of gas to stellar disk scale
length, \emph{the SN distribution is not inconsistent with that of the
ionized gas}.

In reality, the surface brightness distribution of massive young stars in galaxies
is more complex than a simple exponential disk model: indeed, the surface
density of the star
formation rate (SFR) has a hole in the center of our galaxy \citep{Rana+86},
and in other galaxies the SFR follows the surface density of the total (atomic and
molecular) gas \citep{Buat+89,Kennicutt98}, which is known to present a hole
in the central regions \citep[e.g.,][]{Kenn98}.

\begin{table}[h]
\begin{center}
  \caption{Kolmogorov-Smirnov tests of the consistency of different CCSN distributions
with two exponential disk models.
\label{SNvsdisks}}
  \begin{tabular}{lcccc} \hline
    \hline
\multicolumn{1}{c}{Sample}&$\tilde{P}_{\rm KS}^{\rm Freeman}$ &$\hat{P}_{\rm KS}^{\rm Freeman}$& $\tilde{P}_{\rm KS}^{\rm hosts}$&$\hat{P}_{\rm KS}^{\rm hosts}$\\
\multicolumn{1}{c}{(1)}&(2)&(3)&(4)&(5)\\
\hline
Full sample&\textbf{0.002}&\textbf{0.010}&0.895&\textbf{0.010}\\
\hline
    Sa-Sbc&\textbf{0.040}&\textbf{0.028}&0.749&\textbf{0.012}\\
    Sc-Sd&\textbf{0.044}&0.602&0.328&0.166\\
    Sa-Sbc (SNe~Ib/c)&\textbf{0.000}&\textbf{0.016}&\textbf{0.003}&0.091\\
    Sc-Sd \,\ (SNe~Ib/c)&0.111&0.944&0.335&0.973\\
    Sa-Sbc (SNe~II)&0.434&\textbf{0.021}&0.229&\textbf{0.001}\\
    Sc-Sd \,\ (SNe~II)&0.219&0.252&0.693&0.106\\
    Sa-Sd \,\ (without bars)&\textbf{0.006}&0.194&0.302&0.208\\
    Sa-Sd \,\ (with bars)&0.157&\textbf{0.036}&0.814&\textbf{0.013}\\
    Sa-Sbc (without bars)&\textbf{0.037}&0.162&0.703&0.387\\
    Sa-Sbc (with bars)&0.298&0.296&0.390&\textbf{0.006}\\
    Sc-Sd \,\ (without bars)&\textbf{0.036}&0.965&0.428&0.566\\
    Sc-Sd \,\ (with bars)&0.873&0.265&0.686&0.259\\
    $-23 < M_{\rm disk} \leq -20.5$&\textbf{0.003}&0.733&\textbf{0.039}&0.947\\
    $-20.5 < M_{\rm disk} < -18$&0.359&\textbf{0.007}&0.055&\textbf{0.000}\\
    $19.65 < \mu_0^{\rm disk} \leq 20.60$&0.064&0.396&0.940&0.055\\
    $20.60 < \mu_0^{\rm disk} < 21.85$&\textbf{0.023}&\textbf{0.049}&0.676&0.125\\
    $0^\circ \leq i \leq 30^\circ$&\textbf{0.032}&0.889&0.059&0.532\\
    $30^\circ < i \leq 50^\circ$&\textbf{0.029}&\textbf{0.005}&0.797&\textbf{0.003}\\
\hline
    SNe~Ib&\textbf{0.019}&0.302&0.160&0.840\\
    SNe~Ic&\textbf{0.002}&0.109&\textbf{0.042}&0.523\\
    SNe~Ib/c&\textbf{0.000}&0.105&\textbf{0.014}&0.621\\
    SNe~II&0.102&\textbf{0.006}&0.496&\textbf{0.001}\\
    \hline
  \end{tabular}
\end{center}
\noindent Notes: Col.~1: sample;
Col.~2: KS test probabilities that
CCSN subsamples have  isophotal radius - normalized distributions consistent
with the Freeman disk;
Col.~3: same as Col.~2, but with scale-length normalization;
Col.~4: same as Col.~2, but for the host stellar disk;
Col.~5: same as Col.~4, but with scale-length normalization.
CCSN distributions
inconsistent with these exponential models are highlighted in bold.
\end{table}

Table~\ref{SNvsdisks} presents, for the full 224 CCSN sample and
different subsamples, the  KS test probabilities that the
isophotal-radius-normalized
distributions are consistent with those of the stars in a Freeman disk ($\tilde{P}_{\rm
  KS}^{\rm Freeman}$), and of the CCSN host disks ($\tilde{P}_{\rm KS}^{\rm hosts}$).
While there are several departures from the Freeman exponential disk,
most subsamples are consistent with the exponential disk that we
infer for the galaxy hosts. This is most significant for
the full CCSN sample compared to the Freeman disk,
because of the central drop and of the few SNe lying very far out.

\begin{table}[ht]
\begin{center}
  \caption{Kolmogorov-Smirnov tests of the consistency of the
radial distributions of CCSNe among different pairs of subsamples.}
\label{KSsubsamp}
\tabcolsep 5pt
  \begin{tabular}{llcc} \hline
    \hline
\multicolumn{1}{c}{Subsample 1} &\multicolumn{1}{c}{Subsample
  2}&$\tilde{P}_{\rm KS}$&$\hat{P}_{\rm KS}$\\
\multicolumn{1}{c}{(1)} & \multicolumn{1}{c}{(2)} & (3) & (4) \\
    \hline
    Sc-Sd & Sa-Sbc&0.935&0.300\\
    Sa-Sbc (SNe~Ib/c)&Sc-Sd \,\ (SNe~Ib/c)&0.197&0.110\\
    Sc-Sd \,\ (SNe~II)&Sa-Sbc (SNe~II)&0.926&0.268\\
    Sa-Sd \,\ (without bars)&Sa-Sd \,\ (with bars)&0.398&0.214\\
    Sa-Sbc (with bars)&  Sa-Sbc (without bars)&0.711&\,\ 0.313*\\
    Sc-Sd \,\ (without bars)&Sc-Sd \,\ (with bars)&0.294&0.282\\
    $-23 < M_{\rm disk} \leq -20.5$&$-20.5 < M_{\rm disk} < -18$&0.114&\textbf{0.006}\\
    $19.65 < \mu_0^{\rm disk} \leq 20.60$&$20.60 < \mu_0^{\rm disk} < 21.85$&0.469&\,\ 0.723*\\
    $0^\circ \leq i \leq 30^\circ$&$30^\circ < i \leq 50^\circ$&0.273&0.050\\
    $V_{\rm r}\leq5000$&$V_{\rm r}>5000$&\,\ 0.992*&0.568\\
    $V_{\rm r}\leq5000$ (SNe~Ib/c)&$V_{\rm r}>5000$ (SNe~Ib/c)&\,\ 0.444*&\,\ 0.475*\\
    $V_{\rm r}\leq5000$ (SNe~II)&$V_{\rm r}>5000$ (SNe~II)&\,\ 0.924*&0.674\\
    SNe~Ic&SNe~Ib&0.946&\,\ 0.751*\\
    SNe~Ib&SNe~II&0.055&0.156\\
    SNe~Ic&SNe~II&\textbf{0.006}&\textbf{0.035}\\
    SNe~Ib/c&SNe~II&\textbf{0.002}&\textbf{0.015}\\
    \hline
  \end{tabular}
\end{center}
\noindent Notes:
Col.~1: more radially concentrated subsample;
Col.~2: less radially concentrated subsample
(the contrary cases are marked by an asterisks).
CCSN distributions inconsistent within two subsamples are highlighted in bold.
\end{table}

KS tests comparing different subsamples of CCSNe are given in Table~\ref{KSsubsamp}.
These tests confirm our conclusions based upon the maximum likelihood scale
lengths: 1) the surface density distribution of SNe~Ib/c is highly inconsistent with that
of SNe~II, because the former are more centrally concentrated (Table~\ref{hMLall});
2) the distribution of CCSNe is not significantly affected by the morphological
type of the host galaxy, the presence of bars, the luminosity of the host
galaxy or its disk inclination.

\subsection{Scale length normalization}
\label{resscalelennorm}

When we normalize the deprojected radial distances of CCSNe to the indirectly
determined scale lengths of the disks of their host galaxies, we find that
their distribution is even better described by an exponential law.
With the exponential model of
$\Sigma^{\rm SN}(\hat{r})=\Sigma_0^{\rm SN} \exp(-\hat{r}/\hat{h}_{\rm SN})$,
where $\hat{r}=R_{\rm SN}/h$, the maximum likelihood estimate of
$\hat{h}_{\rm SN} = h_{\rm SN}/h$ will satisfy $\hat{h}_{\rm SN} = {1\over 2}\,\left \langle
\hat{r}_i \right \rangle$, as can be found in the same way as
equation~(\ref{htildeML}).
The top right panel of Fig.~\ref{sdensall} shows that the stacked surface density
distribution of all 224 CCSNe appears consistent with an exponential law,
with only a very small drop in the central regions ($R_{\rm SN} < 0.5\,h$).

Table~\ref{hMLall} indicates that all subsamples have surface density profiles that
are consistent with the exponential model. The Table also confirms that
\emph{the mean scale length of SNe~II is 35\% greater than that of SNe~Ib/c,
with the scale length of stars in between}.

The bottom right panel of Fig.~\ref{sdensall} illustrates the different
radial distributions of the SNe~Ib/c and SNe~II.
For both SN types, the drop in the center is much less pronounced
than with the isophotal radius normalization, and not statistically significant.

Table~\ref{SNvsdisks} indicates that nearly all subsamples have
surface densities distributions that
are consistent with the exponential disks of their host galaxies.
The exceptions are early-type, barred, and
low luminosity and highly inclined galaxies, as well as SNe~II
(there are 1 SNe~Ic and 4 SNe~II lying beyond 7 scale lengths).

Table~\ref{KSsubsamp} shows that the distributions of subsamples are consistent when coupled in pairs,
except that \emph{also with scale-length normalization,
SNe~Ib/c are significantly more centrally concentrated than are SNe~II
and that CCSNe are more concentrated in high luminosity host galaxies}.

We also find marginal trends for the CCSNe to be more centrally concentrated
in face-on host galaxies, and no significant difference between
the radial distributions of SNe~Ib and SNe~Ic.

In addition, we performed KS tests for dividing our subsamples of both types of
CCSNe into different distance bins, of the half most nearby and half most distant
host galaxies. These tests confirm that the distribution of CCSNe in our sample
(both isophotal and scale-length normalizations) is not significantly affected
by the distance of the host galaxy (see Table~\ref{KSsubsamp}).

\section{Discussion and Conclusions}
\label{discus}

Our results reported in Sect.~\ref{results} (the top panels of Fig.~\ref{sdensall},
and Table~\ref{hMLall}) indicate
that the global surface density of CCSNe within their host galaxies is exponential
with a scale length of $h_{\rm SN} = 0.29\pm0.01\,R_{25}$, which is 30\%
larger than the scale length found by \cite{vdB97} with a 5 times smaller
sample.

There are several reasons why the observed
radial distribution of CCSNe might not be a perfect exponential:
\begin{enumerate}
\item the total radial distribution may be the combination of several components
  with different distributions;
\item the inner disk may be destroyed or perturbed by an important bulge;
\item the inner CCSNe may be confused with the light of the bulge;
\item dust extinction may prevent the observation of CCSNe in the more opaque
      inner regions of edge-on spiral galaxies;
\item the progenitors of CCSNe may follow a truncated exponential disk.
\end{enumerate}

Indeed, the exponential surface density profile
is the combination of a dominant exponential surface
density profile of SNe~II and a secondary steeper exponential surface density
profile of SNe~Ib/c
(see bottom panels of Fig.~\ref{sdensall}),
so it is more prudent to consider the SN types separately.
For the SNe~II alone, we find a scale length of $0.31\pm0.02\,R_{25}$ which is
consistent with the result of $0.27\pm0.08\,R_{25}$ found by \cite{Bar+92} with a
sample half the size of ours.
But contrary to \citeauthor{Bar+92}, we see no signs of the break in the
exponential slope for SNe~Ib/c at $0.5\,R_{25}$ with
a sample nearly three times as large (bottom left panel of Fig.~\ref{sdensall}).
An exponential model also fits well the surface density
of both types of CCSNe with radii normalized to the
(statistically-determined) disk scale lengths
(bottom right panel of Fig.~\ref{sdensall}).

There is a small loss of CCSNe in
the central regions of host galaxies
(bottom panels of Fig.~\ref{sdensall}).
This is not unexpected since the observed radial distribution of CCSNe can be
affected by the internal dust extinction of their hosts as well as the
confusion with the high surface brightness of their bulges.
In our Galaxy, the star formation rate
as a function of galactocentric radius does not follow a pure exponential
disk, but is vigorous
near the center and is strongly peaked around $R \sim 5 \,\rm kpc$ \citep[e.g.,][]{Kenn89}.
Since the mean normalized
distances of CCSNe in inclined galaxies turn out to be
larger than those in face-on
ones (Tables~\ref{hMLall} and \ref{KSsubsamp}), dust extinction
plays a quantitative role, and should also explain
the central dip in the CCSN surface density profile, mainly seen with the
isophotal normalization
(bottom left panel of Fig.~\ref{sdensall}).
This dip cannot be explained by confusion with the light from the galaxy bulge, since
CCSNe appear more centrally concentrated in the more luminous galaxies.

The dip in the inner distribution of SNe~II has also been recently reported by
\citet{AndeJa09}. These authors found that this central deficit of SNe~II is
offset by a central excess of SNe~Ib/c. Instead, our analysis
(Fig.~\ref{sdensall}) shows no central excess for SNe~Ib/c.  In fact with
the isophotal normalization one sees fairly significant central dips in the
distributions of both SNe~II and SNe~Ib/c, while with the scale length
normalization there is a small dip for the SNe~Ib/c, and only a weak dip at
best for the SNe~II (neither being statistically significant).

Whereas several authors have suggested that stellar disks are sharply
truncated at several scale lengths
(\citealp[e.g.,][]{Ruphy+96} found that the Milky Way is truncated at 6.5 scale lengths),
there are no signs of
such a truncation in the distribution of CCSNe out to
7 scale lengths (right panels of Fig.~\ref{sdensall})
and SNe are observed out to nearly 11 scale lengths.
In fact, with our fairly large sample, we find weak signs of
a shallower slope at large radii, which does not occur at
$1.4\,R_{25}$ as reported by \citeauthor{Bar+92}, but at
$R \simeq 2.0\,R_{25}$ for SNe~II, as we have two SNe~II near $2.4\,R_{25}$
(see bottom left panel of Fig.~\ref{sdensall}).

Since CCSNe are believed to originate from massive stars, one expects
that the radial distribution of CCSNe should resemble that of
tracers of recent star formation
such as ionized gas or perhaps even of
future star formation such as molecular gas.
The values of scale length $\tilde{h}_{\rm SN}$ for CCSNe in spiral host galaxies
(see Col.~3 of Table~\ref{hMLall}) are in good agreement with the data reported by
\citet{Ath+93} on scale lengths of \ion{H}{ii} regions on exponential disks in spiral galaxies of
different types. But the huge uncertainty on the scale length of the ionized gas prevents us
from making a firm conclusion here.

We thus focused our analysis of the
surface density profile of CCSNe by comparing them to the surface density
profile of
stars in spiral disks, which admittedly is a poorer proxy for CCSNe
progenitors than are the \ion{H}{ii} regions.
With the isophotal radius normalization, the stacked surface density profile
of SNe~II is similar to that of the Freeman stellar disk, while the surface
density profile of the SNe~Ib/c has a smaller scale length than that of the
Freeman disk. Given the uncertainties on the relevance of the Freeman disk
for spiral galaxies, we also scaled the SN galactocentric distances to our
statistical estimate of the scale length of each galaxy. Then, we find that
the disk scale length is in between the small scale length of the
SNe~Ib/c and the larger scale length of the SNe~II.

As mentioned above, we also find that SNe~Ib/c
are significantly more concentrated towards the centers of
their host galaxies than are the SNe~II, for the isophotal radius
normalization (bottom left panel of Fig.~\ref{sdensall}, as well as
Table~\ref{KSsubsamp}).
This result has been known for some time
\citep[e.g.,][]{vdB97,Wang+97,Petr+05,Hak08} for the distribution of radii normalized to
$R_{25}$, but we confirm the more centrally concentrated distribution of
SNe~Ib/c with our normalization to the disk scale lengths
(Fig.~\ref{sdensall} and Table~\ref{KSsubsamp}). For both
normalizations, the scale length of the SN~Ib/c distribution is roughly
30\% smaller than that of the SN~II distribution.

We find no statistically significant difference in the radial
distributions of SNe~Ib and SNe~Ic, regardless of the normalization used.
This result may appear in
conflict with \cite{AndeJa09}, who found that SNe~Ic are more centrally
concentrated than SNe~Ib, while our measures of the concentration are
consistent (see values of $\tilde h_{\rm SN}$ and $\hat h_{\rm SN}$ in
Table~\ref{hMLall}).
However \citeauthor{AndeJa09} admit that their result
is not statistically significant. Moreover, the mean distance of SNe~Ic in our
sample is pushed up by one SN~Ic that lies very far from its host galaxy
($1.8\,R_{25}$ and over 10 scale lengths). Still, even if this apparent
outlier is omitted from our sample, the normalized radial distributions of
SNe~Ib and SNe~Ic are similar enough that the KS test fails to distinguish
them with statistical significance. Admittedly, the analysis of
\citeauthor{AndeJa09} had the advantage of a much more precise measurement of the
normalization, as they actually went through the trouble of measuring the
light distribution around each host galaxy.

It is a well established fact that the metallicity
in spiral disks decreases with increasing
galactocentric distance \citep[e.g.,][]{hw99}.
The obvious physical explanation for the more peaked radial distribution of SNe~Ib/c with respect to SNe~II
is the effect of metallicity of SN progenitor environment.
The local metallicity of the SN progenitor environment \citep{BP09} as well
as the global
metallicity of the host galaxy \citep[e.g.,][]{pb03,mm05,PSB08} are approximately correlated
with the SN progenitor metallicity: the ratio of the number of SNe~Ib/c to SNe~II
increases with increasing local metallicity in CCSN hosts.
In this respect, \citet{EMM02} found
that the number ratio of the blue to
red supergiants and the local metallicity in the Milky Way decreases with increasing
galactocentric radius.
\cite{AndeJa09} also concluded that the more concentrated distribution
of SNe~Ib/c must be  related to a strong metallicity dependence on
the relative production of these CCSNe, with SNe~Ib/c arising from higher
metallicity progenitors than SNe~II.
Note that, since higher luminosity spirals have higher metallicity at
the effective radius \citep{Garnett02}, the more concentrated distribution (with scale-length
normalization) of CCSNe in high luminosity galaxies may also be a sign of the
higher metallicity of these hosts.

In summary, our reanalysis of a considerably large sample of CCSNe has allowed us to
derive precise surface density profiles for different subsamples of CCSNe.
These in turn led us to the following statistically based conclusions:
\begin{enumerate}
\item The surface density of CCSNe in all studied subsamples (different CCSN type or galaxy
      host type or luminosity or inclination) falls exponentially with relative
      radius $R_{\rm SN} / R_{25}$ and  $R_{\rm SN} /h$, with no signs of
      truncation out to 7 disk scale lengths;
\item The radial distribution of CCSNe is not significantly affected by the
      host galaxy type, or by the presence of a bar in the host;
\item The radial distribution of CCSNe, measured with scale-length
      normalization, is more concentrated in high luminosity host galaxies;
\item The radial distribution of SNe~Ib/c is significantly more concentrated than the Freeman
      disk distribution and consistent with the radial distribution of \ion{H}{ii} regions;
\item The radial distribution of SNe~II is consistent with both the Freeman
      disk and the \ion{H}{ii} regions distributions, but significantly less
      concentrated than the host disks;
\item There is a small lack of CCSNe
      within one-fifth of the isophotal radius ($R_{\rm SN} < 0.2\,R_{25}$),
      not well visible with scale-length normalization;
\item The radial distribution of type Ib/c SNe in their host galaxies is more centrally
      concentrated than that of type II SNe, the ratio of scale lengths is $0.77\pm0.03$,
      probably because of a metallicity effect.
\end{enumerate}

It would be worthwhile to extend these analyses, by comparing the radial
distribution of CCSNe with recent accurate measures of the distribution of
molecular gas (i.e. CO), as well as ionized gas (see compilation by
\citealp{BPBG03}) and even neutral gas, and by extending the analysis to two
dimensions (extending previous analyses such as those by \citealp{JM63},
\citealp{BTF94}, \citealp{vDHF96}, and \citealp{TBP01}, making use of the
much larger CCSN sample used here).
\begin{acknowledgements}

We thank F. Matteucci, N. Prantzos, and J. Silk for useful discussions.
This research was supported in part by a grant
from the PICS France-Arm\'enie and a scholarship to A.A.H. from the French Government.
A.R.P. wishes to thank the Institut d'Astrophysique de Paris
(France) for the hospitality and support (via the EARA network) during the beginning
stage of this work and the Osservatorio Astrofisico di Catania (Italy) for hospitality during the
last stage of this work.
This research has made use of the Asiago Supernova Catalogue, which is
available at http://web.oapd.inaf.it/supern/cat and the HyperLeda database,
available at http://leda.univ-lyon1.fr/. Finally, we
are especially grateful to our referee for his/her constructive comments.
\end{acknowledgements}

\end{document}